\documentclass[twocolumn,showpacs,preprintnumbers,amsmath,amssymb,superscriptaddress]{revtex4}

\usepackage{graphicx}
\usepackage{epstopdf}
\usepackage{dcolumn}
\usepackage{bm}
\usepackage{color} 

\begin{document}

\title{Magnetic susceptibility of optimally doped HgBa$_{2}$CuO$_{4+\delta}$} 

\author{Yutaka Itoh\thanks{E-mail:yitoh@cc.kyoto-su.ac.jp}}
 \affiliation{Department of Physics, Graduate School of Science, Kyoto Sangyo University, Kamigamo-Motoyama, Kika-ku, Kyoto 603-8555, Japan}
\author{Takato Machi }
 \affiliation{AIST Tsukuba East, Research Institute for Energy Conservation, 1-2-1 Namiki, Tsukuba, Ibaraki 305-8564, Japan}
\author{Ayako Yamamoto}
 \affiliation{Graduate School of Engineering and Science, Shibaura Institute of Technology,\\
  3-7-5 Toyosu, Koto-ku, Tokyo 135-8548, Japan}

\date{\today}

　\begin{abstract}
The magnitude of the powder spin susceptibility of an optimally doped superconductor HgBa$_2$CuO$_{4+\delta}$ (Hg1201) in the normal state is found to be nearly the same as that of La$_{2-x}$Sr$_{x}$CuO$_{4}$ near the optimally doped level.
The Stoner enhancement factor of Hg1201 is larger than that of La$_{2-x}$Sr$_{x}$CuO$_{4}$.
The magnitude correlation of the Stoner enhancement factor is inconsistent with the effect of the recent theoretical Coulomb repulsion between 3$d$ electrons and that of the superexchange intereraction of a charge transfer type.  
\end{abstract}　　

\maketitle
　　　 
\section{Introduction} 
Uniform spin susceptibility $\chi_s$ is a fundamental property of strongly correlated electron systems to understand the many-body effects. 
It is the $\vec{q}$ = 0 component of the static spin susceptibility $\chi^{\prime}_s(\vec{q})$ ($\vec{q}$ is the wave vector). 
For typical high-$T_\mathrm{c}$ cuprate superconductors in the underdoped regime, the normal-state $\chi_s$ decreases with decreasing temperature, which is known to be the pseudogap effect. 
In the optimally doped regime, $\chi_s$ is nearly independent of temperature. 
In the overdoped regime, $\chi_s$ is still nearly independent of temperature, or increases with decreasing temperature, or makes a broad maximum above $T_\mathrm{c}$. 
The magnitude of the temperature independent $\chi_s$ also tells us how different it is from the conventional Fermi liquid theory and how much is the degree of the Coulomb repulsion effect through the Stoner exchange enhancement factor.     
Recent theoretical calculations of the Coulomb repulsion $U$ from first-principles indicate that the lower $T_\mathrm{c}$ La$_{2-x}$Sr$_x$CuO$_4$ has stronger $U$ than HgBa$_{2}$CuO$_{4+\delta}$ (Hg1201) by 1.47 times~\cite{Kuroki}. 
The estimation of the Stoner factor could be an experimental test of the Coulomb repulsion $U$ in the itinerant systems.  

In this paper, we report on the measurement and analysis of the bulk magnetic susceptibility of a single-CuO$_2$-layer high-$T_\mathrm{c}$ superconductor Hg1201 at the optimally doped level ($T_\mathrm{c}$ = 98 K) in the normal state.    
We found that the Stoner enhancement factor of Hg1201 is larger than that of La$_{2-x}$Sr$_{x}$CuO$_{4}$, which is in contrast to the recent theoretical calculations on $U$. 
Discussions were made from a single band Hubbard model and a $t$-$J$ model within the random phase approximation. 

\section{Experiments}
High quality polycrystalline samples of Hg1201 were prepared by a solid state reaction with high purity BaO powder, 
which was a key to heat treatment at a relatively high temperature of 930 $^{\circ}$C~\cite{Yamamoto}. 
The sample for the present study has been confirmed to be in single phase by powder X-ray diffraction patterns and characterized by transport measurements~\cite{Yamamoto}.   
The dc magnetic susceptibility $\chi$ at an external magnetic field of 1.0 T was measured by a superconducting quantum interference device (SQUID) magnetometer (QUANTUM Design, MPMS). 
From the previous magnetization measurement in a field of 20 Oe, the superconducting transition temperature $T_\mathrm{c}$ was estimated to be 98 K~\cite{Yamamoto}. 
For the comparative studies on the powder spin susceptibility, the polycrystalline samples of La$_{2-x}$Sr$_x$CuO$_4$ ($x$ = 0.13, 0.15 and 0.18) were synthesized by a conventional solid stat reaction method~\cite{LSCO}. After they were confirmed to be of a single phase by the powder X-ray diffraction patterns, their powder magnetic susceptibilities were measured by the SQUID magnetometer. 
The samples of $x$ = 0.13, 0.15, 0.18 exhibit $T_\mathrm{c}$ = 34, 38, 35 K, respectively. 

The powder susceptibility $\chi$ is the isotropic part of the magnetic susceptibility ($\chi_{aa} + \chi_{bb} + \chi_{cc}$)/3
($\chi_{\alpha\alpha}$ in a field along $\alpha$ axis).   
The anisotropic part of the magnetic susceptibility is known to be due to the Van Vleck orbital susceptibility~\cite{Shimizu,Walstedt}. 
Although the recently grown single crystals~\cite{Greven} will enable us to obtain the tensor components in $\chi_{\alpha\alpha}$, 
the powder magnetic susceptibility can tell us the isotropic part of the spin susceptibility and how much is the electron correlation effect. 

\section{Experimental results}
\subsection{spin susceptibility of Hg1201}
Figure 1 shows the magnetic susceptibility $\chi$ of the powder Hg1201. 
$\chi$ levels off above about 250 K and decreases with cooling to $T_\mathrm{c}$ = 98 K,
which is a pseudogap behavior. 
Since the temperature dependence of $\chi$ resembles those of the plane-site $^{63}$Cu and $^{17}$O Knight shifts~\cite{Bob,Itoh1,Haase},
one may regard the present bulk magnetic susceptibility $\chi$ as an intrinsic behavior. 
\begin{figure}[t]
 \begin{center}
 \includegraphics[width=1.08\linewidth]{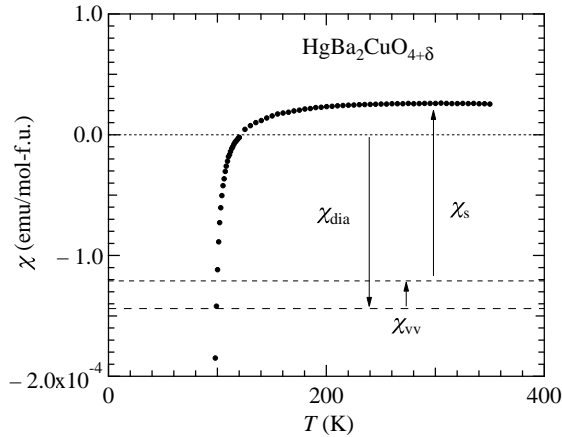}
 \end{center}
 \caption{\label{f2}
Bulk magnetic susceptibility $\chi_\mathrm{}$ of powdered HgBa$_{2}$CuO$_{4+\delta}$ ($T_\mathrm{c}$ $\approx$ 98 K) in an external magnetic field of 1.0 T. The bulk $\chi_\mathrm{}$ is the sum of inner core electron diamagnetic susceptibility $\chi_\mathrm{dia}$($<$ 0), Van Vleck orbital susceptibility $\chi_\mathrm{vv}$($>$ 0), and 3$d$ electron spin susceptibility $\chi_\mathrm{s}$.  
 }
 \end{figure}

\begin{figure}[t]
 \begin{center}
 \includegraphics[width=0.89\linewidth]{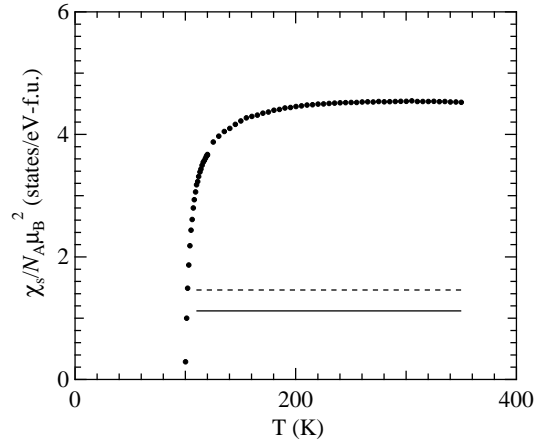}
 \end{center}
 \caption{\label{f3}
Uniform spin susceptibility $\chi_s$/$N_\mathrm{A}\mu_\mathrm{B}^2$ in states/eV-f.u. for HgBa$_{2}$CuO$_{4+\delta}$ ($T_\mathrm{c}$ $\approx$ 98 K). 
Solid and dashed lines are the electron density of states at the moderately doped level and the two dimensional van Hove singularity from the band theoretical calculations~\cite{BandHg1201}. 
 }
 \end{figure} 

The bulk magnetic susceptibility is the sum of the individual components, 
$\chi = \chi_s +\chi_\mathrm{vv}+\chi_\mathrm{dia}$, 
where $\chi_{\mathrm{s}}$ is the spin susceptibility of the 3$d$ electrons, 
$\chi_{\mathrm{vv}}$ is the powder-averaged Van Vleck orbital susceptibility (+0.23$\times$10$^{-4}$ emu/mole-f.u. after the band calculation for Sc$_2$CuO$_4$~\cite{SCO} and according to the analysis for La$_{2-x}$Sr$_{x} $CuO$_4$~\cite{Walstedt,Johnston}), and 
$\chi_{\mathrm{dia}}$ is the diamagnetic susceptibility of the inner shell electrons in the core ($-$1.44 $\times$ 10$^{-4}$ emu/mole-f.u.~\cite{dia}).  
The spin susceptibility $\chi_s$ is obtained from $\chi_s = \chi-\chi_\mathrm{vv}-\chi_\mathrm{dia}$. 
The estimated $\chi_s$ = 1.47 $\times$ 10$^{-4}$ emu/mole-f.u. of Hg1201 ($T >$ 250 K) in Fig. 1 is nearly the same as the reported $\chi_s$ = 1.4 $-$ 1.7 $\times$ 10$^{-4}$ emu/mole-f.u. of La$_{2-x}$Sr$_{x}$CuO$_{4}$ near the optimally doped level~\cite{Cava,Terasaki,Johnston,Hucker}. 

Figure 2 shows the spin susceptibility $\chi_s/N_\mathrm{A}\mu_\mathrm{B}^2$ of an optimally doped Hg1201, where $N_\mathrm{A}$ is the Avogardro's number and $\mu_{B}$ is the Bohr magneton. 
Solid and dashed lines indicate the bare spin susceptibility $\chi_0/N_\mathrm{A}\mu_\mathrm{B}^2$ = $N(E_\mathrm{F})$ at the moderately doped level and the two dimensional van Hove singularity from the band theoretical calculation~\cite{BandHg1201}, 
where $N(E_\mathrm{F})$ is the electron density of states at the Fermi energy $E_\mathrm{F}$ for both spin directions in units of states/eV-f.u. and the electron $g$-factor is assumed to be 2. 

\subsection{Hg1201 vs La$_{2-x}$Sr$_x$CuO$_4$}
Figure 3(a) shows the powder magnetic susceptibilities $\chi$ in emu/g of La$_{2-x}$Sr$_x$CuO$_4$ ($x$ = 0.13, 0.15, 0.18) in an external magnetic field of 1.0 T for comparison with Hg1201.   
The magnitude of $\chi$ of La$_{2-x}$Sr$_x$CuO$_4$ is nearly the same as those in the literatures~\cite{Cava,Terasaki,Johnston,Hucker}.

Figure 3(b) shows the powder spin susceptibilities $\chi_s$ in emu/mol-f.u. of La$_{2-x}$Sr$_x$CuO$_4$, which are obtained from $\chi_s = \chi-\chi_\mathrm{vv}-\chi_\mathrm{dia}$ in the same procedure of the spin-orbital partition as Hg1201.
The orbital susceptibility $\chi_{\mathrm{vv}}$ is taken to be +0.23$\times$10$^{-4}$ emu/mole-f.u.~\cite{Walstedt,SCO,Johnston}.  
The core electron susceptibility $\chi_{\mathrm{dia}}$ with the Sr concentration $x$ is taken to be ($-$99+5$x$)$\times$10$^{-6}$ emu/mole-f.u.~\cite{dia}.  

 \begin{figure}[h]
 \begin{center}
 \includegraphics[width=1.05\linewidth]{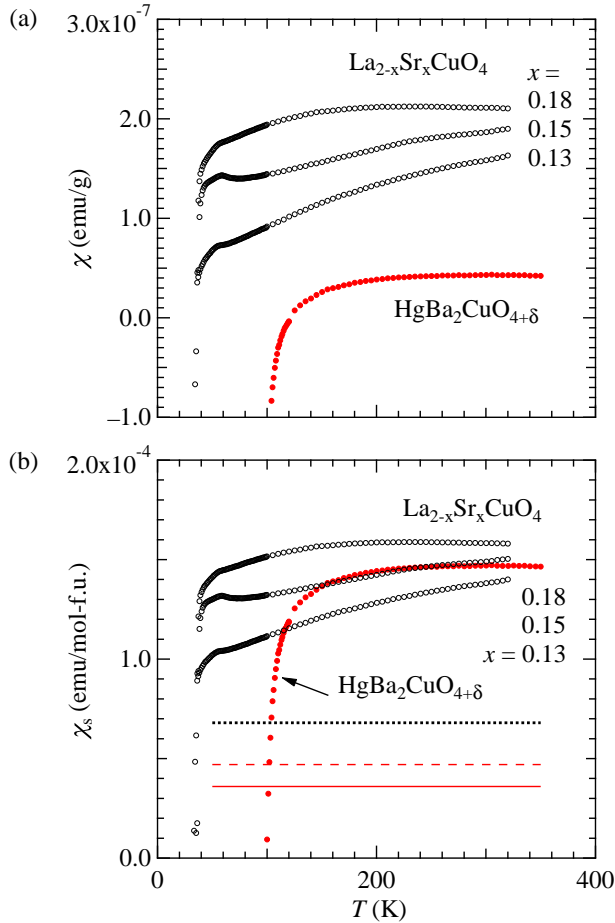}
 \end{center}
 \caption{\label{f3}(Color online)
(a) Powder magnetic susceptibilities $\chi_\mathrm{}$ in emu/g of optimally doped Hg1201 and La$_{2-x}$Sr$_x$CuO$_4$ ($x$ = 0.13, 0.15, 0.18) in an external magnetic field of 1.0 T. 
(b) Powder spin susceptibilities $\chi_s$ in emu/mol-f.u. of optimally doped Hg1201 and La$_{2-x}$Sr$_x$CuO$_4$ ($x$ = 0.13, 0.15, 0.18).  
Dotted line is the band theoretical spin susceptibility $\chi_0$ for La$_{2-x}$Sr$_x$CuO$_4$~\cite{BandLSCO,vHsLSCO}.   
Solid and dashed lines are the band theoretical spin susceptibility $\chi_0$ for Hg1201~\cite{BandHg1201}.
 }
 \end{figure} 
In Fig. 3(b), the magnitude of the spin susceptibility $\chi_s$ of Hg1201 above 250 K is nearly the same as those of La$_{2-x}$Sr$_x$CuO$_4$ for $x$ = 0.13, 0.15, 0.18. 
The band theoretical calculations of the spin susceptibility $\chi_0$ are shown by a dotted line for La$_{2-x}$Sr$_x$CuO$_4$~\cite{BandLSCO,vHsLSCO}
and by solid and dashed lines for Hg1201~\cite{BandHg1201}.

The band theoretical $N(E_\mathrm{F})$ of La$_{1.85}$Sr$_{0.15}$CuO$_4$~\cite{BandLSCO,vHsLSCO} is larger than that of Hg1201~\cite{BandHg1201} by about 1.4 $-$ 1.8 times, while the experimental $\chi_s$ of La$_{1.85}$Sr$_{0.15}$CuO$_4$ is estimated to be nearly the same as that of Hg1201.
For Hg1201, we estimated the Stoner exchange enhancement factor $S$ = $\chi_s$/$\chi_0$ [= 1/(1$-$$IN(E_\mathrm{F})$) in the random phase approximation for an effective interaction $I$] to be 4.1 at the moderately doped level and 3.2 at the two dimensional van Hove singularity. 
Smaller orbital susceptibility $\chi_\mathrm{vv}$ = +0.15$\times$10$^{-4}$ emu/mole-f.u. estimated experimentally in~\cite{Hucker} leads to more enhanced $S$ in $\chi_s$. 
The Stoner enhancement factor $S$ of Hg1201 is 1.5 $-$ 2.0 times larger than $S$ $\sim$ 2 of La$_{1.8}$Sr$_{0.2}$CuO$_4$~\cite{Johnston} and $S$ = 2.0$-$2.3 (320 K) of the present La$_{2-x}$Sr$_{x}$CuO$_4$ ($x$ = 0.13, 0.15, 0.18) in Fig. 3.(b).
The effective interaction $I$ of Hg1201 is stronger than that of La$_{2-x}$Sr$_{x}$CuO$_4$, $I$(Hg1201) $>$ $I$(LSCO).
The estimated parameters are shown in Table I.    
\begin{table}
\caption{Electron density of states $N(E_\mathrm{F})$'s from band theoretical calculations~\cite{BandLSCO,vHsLSCO,BandHg1201}, experimental spin susceptibilities $\chi_s$'s for La$_{1.8}$Sr$_{0.2}$CuO$_4$~\cite{Johnston} and the present Hg1201 ($T >$ 250 K), Stoner enhancement factors $S$'s, and effective interactions $I$'s. 
$N(E_\mathrm{F})$ and $\chi_s/N_\mathrm{A}\mu_\mathrm{B}^2$ are shown in state/eV formula units, and $I$ in eV.  
}
\label{t1}
\begin{center}
\begin{tabular}{lccrc}
\hline
\multicolumn{1}{l}{ } & \multicolumn{1}{r}{$N(E_\mathrm{F})$} & \multicolumn{1}{c}{~$\chi_s/N_\mathrm{A}\mu_\mathrm{B}^2$~}  & \multicolumn{1}{c}{$S$} & \multicolumn{1}{c}{~$I$} \\
\hline
La$_{1.8}$Sr$_{0.2}$CuO$_4$ & 2.09 & 4.3 & 2.1 &~ 0.25 \\
HgBa$_2$CuO$_{4+\delta}$ & 1.46 & 4.5 & 3.2 &~ 0.47 \\
  & 1.12 & 4.5 & 4.1 &~ 0.68 \\ 
\hline
\end{tabular}
\end{center}
\end{table}

\section{Discussions}
The recent first-principles calculations indicate that the on-site Coulomb repulsion $U$ between 3$d$ electrons in Hg1201 is weaker than that in La$_{2-x}$Sr$_{x}$CuO$_4$ by 0.68 times~\cite{Kuroki}.
The value of $U$ = 3.15 eV in La$_{2}$CuO$_4$ is calculated to be larger than $U$ = 2.15 eV in Hg1201~\cite{Kuroki}. 
In the two dimensional Hubbard model with the random-phase approximation, the effective interaction $I$ in the Stoner factor is an effective Coulomb repulsion $\bar{U}$~\cite{BS},
\begin{equation}
\chi_s = {\chi_0\over {1 - \bar{U}N(E_\mathrm{F})}}.
\label{}
\end{equation}
The band theories indicate $N(E_\mathrm{F})$ of Hg1201 smaller than that of La$_{2-x}$Sr$_{x}$CuO$_4$ in Table I~\cite{BandLSCO,vHsLSCO,BandHg1201}.
If one assumes that the magnitude correlation on $\bar{U}$ between Hg1201 and La$_{2-x}$Sr$_{x}$CuO$_4$ (LSCO) is the same as $U$,
that is $\bar{U}$(Hg1201) $<$ $\bar{U}$(LSCO),
the Stoner enhancement factor of Hg1201 must be smaller than that of La$_{2-x}$Sr$_{x}$CuO$_4$.
The present experimental estimation of the Stoner enhancement factor of Hg1201 larger than that of La$_{2-x}$Sr$_{x}$CuO$_4$ is in contrast to the effect of the theoretical Coulomb repulsion $U$. 

In the two dimensional $t$-$J$ model with the random-phase approximation, the effective interaction $I$ in the Stoner factor corresponds to a superexchange interaction $J_{0}$~\cite{TKF}, 
\begin{equation}
\chi_s = {\chi_0\over {1 + J_{0}N(E_\mathrm{F})}}.
\label{}
\end{equation}
The superexchange interaction $J_{0}$ of a charge transfer type is expressed as
\begin{equation}
J_{0} \propto {4T_{pd\sigma}^4\over {\Delta_{ct}^2}}({1\over {\Delta_{ct}}}+{1\over {U}}),
\label{}
\end{equation}
where $T_{pd\sigma}$ is a $p$-$d$ hybridization matrix element and $\Delta_{ct}$ is a charge transfer gap~\cite{ZSA}.  
The value of $T_{pd\sigma}$ in La$_{2}$CuO$_4$ is nearly the same as that in Hg1201~\cite{Kuroki}.
The value of the $d$-$p$ charge transfer energy $\Delta_{dp}$ = 2.58 eV in La$_{2}$CuO$_4$ is larger than $\Delta_{dp}$ = 1.84 eV in Hg1201~\cite{Kuroki}. 
According to Eq.~(3), larger $U$ and $\Delta_{ct}$($\propto\Delta_{dp}$) in La$_{2-x}$Sr$_{x}$CuO$_4$ than Hg1201 lead to 
smaller $J_{0}$ in La$_{2-x}$Sr$_{x}$CuO$_4$ than Hg1201 [$J_{0}$(Hg1201) $>$ $J_{0}$(LSCO)]. 
The Stoner enhancement factor in Eq.~(2) for Hg1201 is smaller than that for La$_{2-x}$Sr$_{x}$CuO$_4$,
which is also in contrast to the experimental magnitude correlation in the Stoner factor. 
No parent antiferromagnetic insulator has been found in Hg1201, maybe due to chemical instability.  
It could not be tested whether the magnitude of $J_0$ is larger in the Mott insulating state of Hg1201 than that of La$_2$CuO$_4$. 

The single-CuO$_2$-layer superconductor Tl$_2$Ba$_2$CuO$_{6+\delta}$ ($T_\mathrm{c}$ = 85 K) also shows a large Stoner enhancement factor $S$ = 3.8$-$4.2 $\sim$ 4~\cite{TL2201}, which is comparable to the present Hg1201 ($T_\mathrm{c}$ = 98 K) in Table I.
The large Stoner enhancement factor may characterize the higher $T_\mathrm{c}$ superconductors than La$_{2-x}$Sr$_{x}$CuO$_4$ and Bi$_2$Sr$_2$CuO$_{6+\delta}$~\cite{Bi2201}. 
 
\section{Conclusions}
In conclusion, the magnitude of the uniform spin susceptibility of the optimally doped Hg1201 is nearly the same as that of La$_{2-x}$Sr$_x$CuO$_4$. 
The Stoner enhancement factor of Hg1201 is larger than that of La$_{2-x}$Sr$_x$CuO$_4$. 
The effective interaction $I$ is in contrast to the recent first-principles calculations on the Coulomb repulsion $U$
and the effect of the superexchange interaction $J_{0}$ in the $t$-$J$ model within the random phase approximation.


\end{document}